\documentclass[useAMS,usenatbib]{mn2e}
\usepackage{graphicx}
\usepackage{epstopdf}
\usepackage[figuresright]{rotating}
\usepackage{subfigure,color}
\usepackage{longtable}
\usepackage{xspace}

\newcommand{\ion}[2]{{\textrm{#1}}\,{\textrm{\sc #2}}}

\definecolor{pink}{rgb}{.9,.2,.5}  
\definecolor{purple}{rgb}{.5,.6,.7}

\title[Nitrogen  abundances  in AGN]
{New quantitative  nitrogen abundance estimations  in  a sample of Seyfert 2 Active Galactic Nuclei}

\author[Dors et al.]
            { 
              O.~L.\ Dors Jr.$^{1}$\thanks{E-mail:olidors@univap.br}, 
	     K.~Z. Arellano-C\'ordova$^{2}$,
              M.~V. Cardaci$^{3,4}$, 
              G.~F.\ H\"agele$^{3,4}$, \\
$^1$ Universidade do Vale do Para\'iba, Av. Shishima Hifumi, 2911, Cep
12244-000, S\~ao Jos\'e dos Campos, SP, Brazil\\ 
$^2$ Instituto Nacional de Astrof\'isica, \'Optica y Electr\'onica (INAOE),
Apdo. Postal 51 y 216, Puebla, Mexico \\
$^3$ Instituto de Astrof\'isica de La Plata (CONICET-UNLP), Argentina. \\
$^4$ Facultad de Ciencias Astron\'omicas y Geof\'{\i}sicas, Universidad Nacional de La Plata, Paseo del Bosque s/n, 1900 La Plata, Argentina.\\} 

\begin{document}

\date{Accepted 2015 Month  00. Received 2015 Month 00; in original form 2014 December 17}

\pagerange{\pageref{firstpage}--\pageref{lastpage}} \pubyear{2011}

\maketitle

\label{firstpage}

\begin{abstract}
 
We obtained  new  quantitative  determinations  of the nitrogen abundance  and a consistent relation between nitrogen 
and oxygen abundances for a sample of Seyfert 2 galaxies located at redshift $z < 0.1$. We carried out this analysis using 
the {\sc Cloudy} code to build detailed photoionization models.  We were able to reproduce  observed optical narrow 
emission line intensities for 44 sources compiled from the literature. 
 Our results show that Seyfert 2 nuclei have nitrogen abundances ranging from $\sim0.3$ to $\sim 7.5$ times the solar value. 
 We derived the relation $\rm \log(N/H)=1.05 (\pm0.09) \times [\log(O/H)] -0.35 (\pm 0.33$). Results for N/O vs. O/H 
   abundance ratios derived for Seyfert 2 galaxies are in consonance with those  recently derived for a sample of 
  extragalactic disk \ion{H}{ii} regions with high metallicity.   
\end{abstract}



\section{Introduction}
The spectra of Active Galactic Nuclei (AGNs) present 
strong emission lines of heavy elements that are easily measured, 
even in objects at  high redshifts. The line intensities can be used to estimate the metallicity of these objects. 
Therefore, AGN  metallicity determinations play a fundamental role in the knowledge of the 
 chemical evolution of  galactic nuclei (e.g. \citealt{dors14,  matsuoka09,
nagao06}). They also provide an indirect understanding  of the star formation history  for the central  parts of galaxies 
\citep{hamann92, hamann93, collin99, wang11}.

The first abundance determinations based on optical emission lines  in  AGNs  were carried out by \citet{osterbrock75} for the
radio galaxy 3C\,405 (Cygnus A).   These authors calculated the abundances of helium and   of  other heavy elements 
(relative to the hydrogen abundance) through determinations
of the electron temperature  of  the AGN gas, i.e. using the $T_{\rm e}-\rm method$ (e.g. \citealt{peimbert69}).
Osterbrock and collaborators, with the goal  of  increasing the sample,   
produced a spectrophotometric survey of radio galaxies   
obtaining the physical conditions of the nuclear gas of several objects. Results of this analysis were published  
by \citet{costero77} and  \citet{koski78}. After these pioneer works, several authors have addressed efforts in order to
determine  chemical abundances in AGNs (see \citealt{hamann99} and references therein).
 However, most of these works have considered only the metallicity parametrized through the oxygen abundance determinations (e.g. \citealt{dors15, dors14,
richardson14, matsuoka09, nagao06, groves06, thaisa98, cruz91}).

In particular, the knowledge of the nitrogen abundances  in ionized nebulae and AGNs is essential for the study 
 of chemical evolution of galaxies, stellar  nucleosynthesis and   stellar material ejection  in the  interstellar medium.  This  also helps to  verify  the presence of 
Wolf-Rayet stars in metal rich environments.
Nitrogen abundances are well-known in star-forming regions. In fact, studies based on spectroscopic data of  \ion{H}{ii} regions
 have shown that this element has  a primary origin for  the low metallicity  regime ($\rm [12+log(O/H)] \la 8.2$)
 and   a  secondary  one  for the
 high metallicity regime (e.g. \citealt{edmunds90, perez09, pily10}). 
However,  nitrogen abundances are poorly known in  AGNs and most abundance determinations
of this element in these objects are qualitative. \citet{thaisa90}  compared the intensity of the [N\,{\sc ii}]($\lambda$$\lambda$6548,6584)/H$\alpha$ and
  [S\,{\sc ii}]($\lambda$$\lambda$6716,31)/H$\alpha$ 
 ratios predicted by photoionization models with observational data from a sample of 177 Seyfert 2   galaxies  (hereafter  Sy2s). These authors found that
models assuming nitrogen abundances ranging from 0.5 to 3 times the solar values reproduce the observational
data  (see also \citealt{yu11, bradley04}).  Moreover, the majority of
the photoionization models built in order to reproduce AGN emission line intensities  
assumed a N/O-O/H relation derived from \ion{H}{ii} region abundance determinations
\citep{dors15, dors14, matsuoka09, groves06, nagao06}.  If these relations are not
valid for AGN, incorrect metallicity determinations or ca\-li\-bra\-ti\-ons between metallicity 
and strong emission-lines  could be obtained \citep{perez09}.

In this letter, we  present   a new quantitative nitrogen abundance determinations (relative to   the  hydrogen abundances) for a 
 sample of  Sy2s,  making also possible to obtain  the N/H-O/H relation
 for these objects.  For this purpose, we compiled optical narrow emission line intensities from the literature  and 
 we built detailed photoionization models to reproduce the observational data of each object of the sample. 
In Section~\ref{dat} we describe the observational data. In Section~\ref{mod} we present the 
photoionization model description and the methodology used to fit the detailed models
to the observational data. In Sections~\ref{resdisc} and~\ref{conc} we present our results and conclusions, respectively.
 
\section{AGN sample}
\label{dat}
We compiled from the literature narrow emission  line intensities of AGNs  classified as Seyfert 2 and 1.9
observed in the optical  range ($\rm 3000 \: \AA  \: < \: \lambda \: < \: 7000 \: \AA$).  
In general,  Seyfert 1 nuclei present a secondary source of ionization and heating, 
i.e.  shocks with high velocity,  not considered by the {\sc Cloudy} code.
Therefore, these objects were not considered in our sample.

We established as selection criterion the presence of the  
 [\ion{O}{ii}]$\lambda$$\lambda$3726+29 (hereafter [\ion{O}{ii}]$\lambda$3727),  [\ion{O}{iii}]$\lambda$5007, [\ion{N}{ii}]$\lambda$6584 and
  [\ion{S}{ii}]$\lambda\lambda$6716+31
narrow emission lines with full width at half-maximum (FWHM) lower than 1000 km/s.
The sample consists of  47  Sy2  nuclei  compiled by \citet{dors15} and 14 observed by \citet{dopita15}.
All emission line intensities are  reddening corrected. 
The AGN sample is composed of objects with redshift $z \la 0.1$,  
observed with long-slit spectroscopy and with integral-field spectroscopy, i.e.
an heterogeneous sample. \citet{castro17} and \citet{dors15} showed that no bias
is introduced in chemical abundance studies if heterogeneous sample of data is
considered. In Table~\ref{tab1}, we list the object identification, the observational 
and model predicted emission line intensities (relative to H$\beta$=1.0),  redshift values taken from NED
\footnote{The NASA/IPAC Extragalactic Database ({\sc ned}) is operated by the Jet 
Propulsion Laboratory, California Institute of Technology, under contract with the National Aeronautics and 
Space Administration.}, and the references from which the data were taken. 

\section{Photoionization model}
\label{mod}
\subsection{Initial parameters}
\label{mod1}
 
We used the {\sc Cloudy} code version 13.04 \citep{ferland13} to build individual photoionization models 
in order to reproduce the observed emission line intensities for each object of our sample.  For each object, we built a first model assuming the following  initial parameters:
\begin{enumerate}
\item Number of ionizing photons [$Q(\rm H)$]-- The logarithm of the number 
of ionizing photons was considered to be equal to 51 dex, 
a typical value  derived for Seyfert galaxies \citep{riffel09}.
 
\item Spectral Energy Distribution  (SED)-- The SED of the ionizing source was modelled as a power law of the 
form  $F_{\nu} \sim \nu^{\alpha}$, where $\alpha$ was assumed to be  equal to $\alpha=-1.4$, a typical value for AGNs (e.g.  \citealt{zamorani81}).

\item Electron density ($N_{\rm e}$)--  We assume a constant value  for the electron density $N_{\rm e}$ along the radius of the hypothetical AGN  emission region. 
This value was derived, for each object,  by using the observed  [\ion{S}{ii}]$\lambda$6716/$\lambda$6731 ratio, obtained from the original work from 
which the observational data were taken (see Table~\ref{tab1}),
and using its relation with $N_{\rm e}$ obtained by \citet{hagele08}.

\item Inner and outer radius-- The inner radius ($R_{\rm in}$), defined as being  the distance from the ionizing source to the illuminated 
gas region, was considered to be equal to 3 pc,  a typical value for narrow line regions of Seyfert galaxies (e.g. \citealt{balmaverde}). 
 The outer radius was assumed to be the one where the electron temperature of the gas reaches 4\,000 K, 
 the default lowest-allowed kinetic temperature by the {\sc Cloudy} code. Gas cooler than $\sim$4\,000 K
 does practically not emit optical emission lines.

\item Metallicity ($Z$)-- For each model,  we assume an initial value of  $Z$ obtained from the calibration proposed by  \citet{castro17}: 
\begin{eqnarray*}
\rm 
(Z/Z_{\odot})\!\!\!&=&\!\!\!1.08(\pm0.19) \times x^2  +  1.78(\pm0.07) \times x +1.24(\pm0.01),   
\end{eqnarray*}
 being $x=\rm \log \left( [N\:II]\lambda6584/ [O\: II]\lambda3727 \right)$  calculated from the observational data listed in Table~\ref{tab1}.

\item Nitrogen and sulphur abundances-- The   nitrogen (N/H) and sulphur (S/H)  abundances relative  to the hydrogen abundance  
 were considered to be  2 and 1 times the solar values, respectively. In the {\sc Cloudy} code  the  solar values 
  12+log(N/H)$_\odot$=7.93 and 12+log(S/H)$_{\odot}=7.27$ were taken from  \citet{holweger01} and  \citet{grevesse98}, respectively.
\end{enumerate}

\begin{figure*}
\centering
\includegraphics[angle=-90,width=0.8\columnwidth]{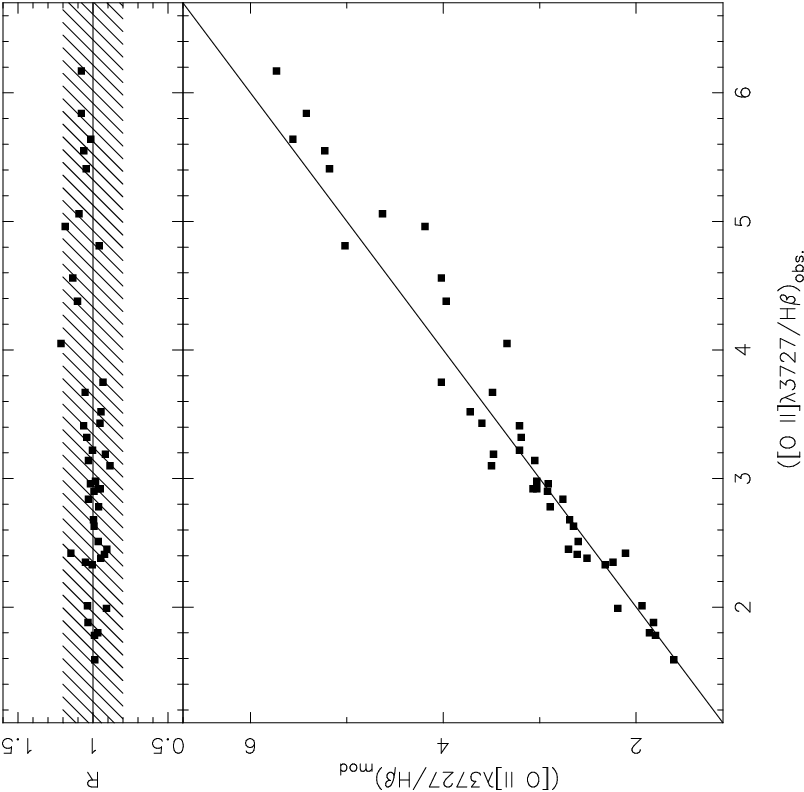}\hspace{0.4cm}
\includegraphics[angle=-90,width=0.8\columnwidth]{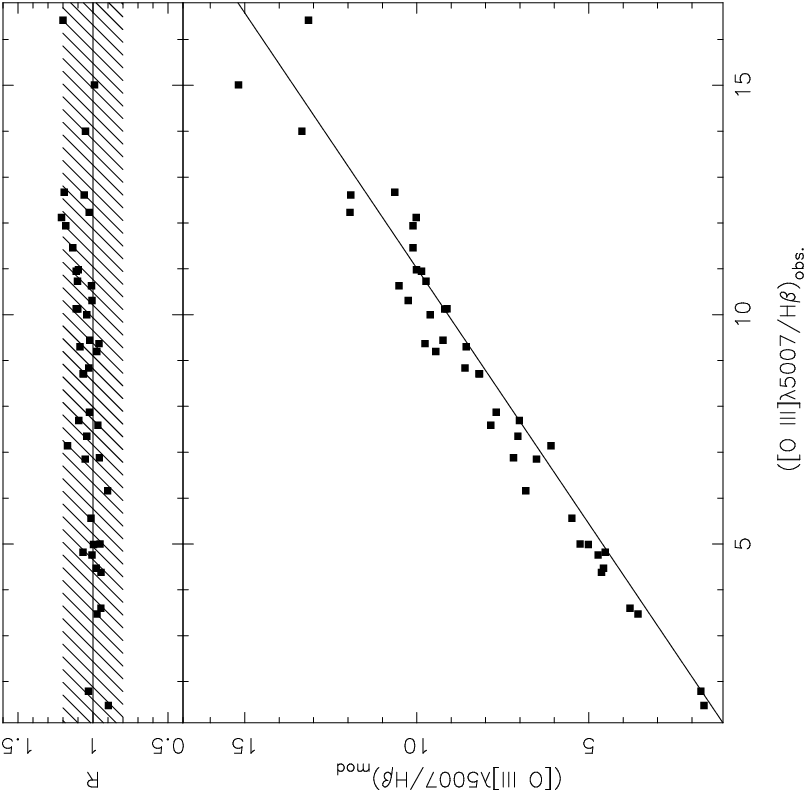}\\\vspace{0.4cm}
\includegraphics[angle=-90,width=0.8\columnwidth]{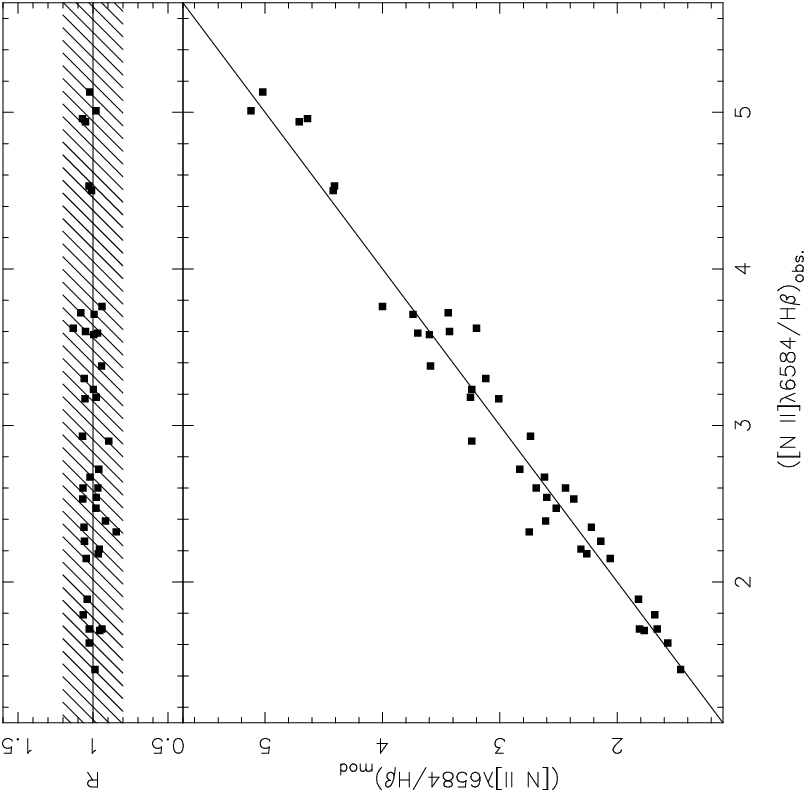}\hspace{0.4cm}
\includegraphics[angle=-90,width=0.8\columnwidth]{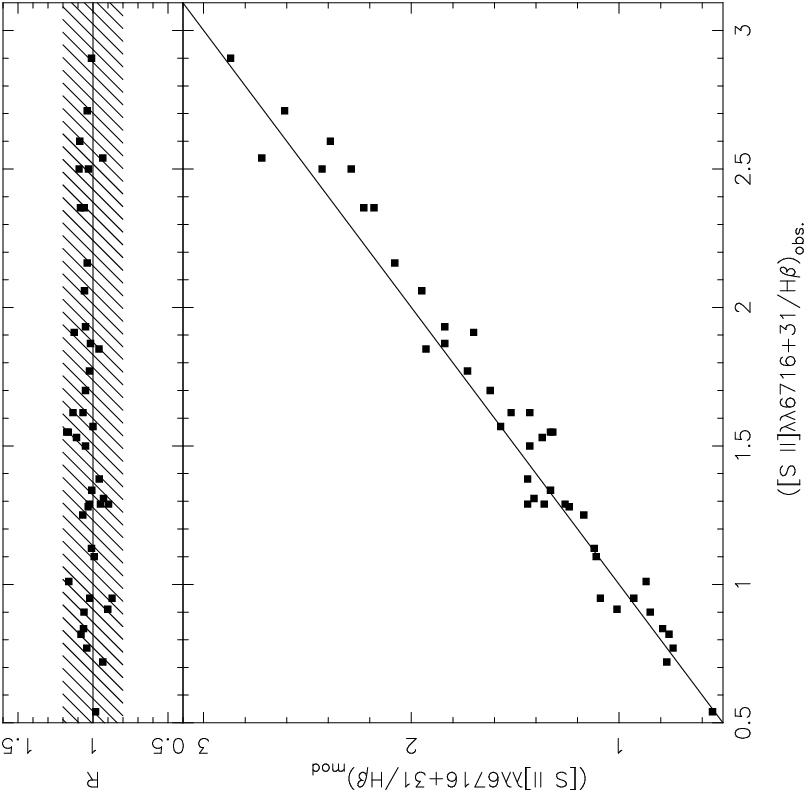}\\\vspace{0.4cm}
 \caption{Bottom  part of each panel: comparison between model predicted and observed emission line intensities 
  (with respect to H$\beta$) for the  sample of Sy2s (see Sect~\ref{resdisc}). Solid lines represent equality of the two estimates.
 Top  part of each panel: ratio between the observed and predicted  intensities of a given emission line versus the observed emission line intensity  for the same ratio.
 The hatched area indicates the band of $20 \%$  adopted for the agreement between the 
 observed and predicted   emission line intensities (see  Sect.~\ref{mod2}). }
\label{fres}
\end{figure*}

\subsection{Fitting model methodology}
\label{mod2}

 An initial model assuming the input parameters described  in Section \ref{mod1} was built
 in order to reproduce the emission line intensities of each object of the sample.
Then, we  ran new models varying separately,  the $Z$, N/H, S/H values considering a step of $\pm$0.2 dex,   typical uncertainty in  
nebular abundance estimations derived through  photoionization models \citep{dors11}.   
 From this series of models, we selected  one that best  
reproduces the  intensities of all emission  lines considered (see Sect.~\ref{dat})  within an uncertainty of $\pm 20 \%$,  which is a typical observational 
uncertainty for emission lines  (e.g. \citealt{kraemer94}).  If no model was able to reach this criterion,
a new series of models  was built varying   $N_{\rm e}$,  $R_{\rm in}$      and $\alpha$  
with a step of $\pm$0.2 dex.  Only one parameter was varied  at a  time and the   
optimization method {\sc phymir} \citep{vanhoof97}  was considered to  select the best fitting  model
to the set of emission line intensities. Similar methodology was adopted by \citet{dors11} in a study of chemical 
abundances in \ion{H}{ii} regions.

To determine the error in our abundance estimations, a simple model simulation
was performed. Initially, we built a photonionization
model  assuming solar  metallicity, $N_{\rm e}$=500 cm$^{-3}$ (typical
electron density value derived  for AGNs, see \citealt{dors15}), and the initial parameters
listed above. Thereafter,  a series of models   varying 
 the O/H, N/H and S/H abundances by a factor of   $\pm$0.5 dex  (step of 0.05 dex) from the solar values   was built.
Using these models, we obtained an abundance range 
 for which emission line intensities differ $\pm20 \%$ from the values predicted by the initial model. We found that a variation of about $\pm$0.1 dex in   abundances
  yields   variations of about 20\% in the intensities of the emission lines considered. This value will be considered as the  uncertainty in our abundance estimations.

\section{Results and Discussion}
\label{resdisc}

\begin{table*}\small
\caption{Dereddened fluxes (relative to H$\beta$=1.00)  for  a sample of Seyfert 2 nuclei.
The observed values compiled from the literature are referred as "Obs." while the   predicted values
by the photoionization models as  "Mod." (see Sect.~3 of the Letter).  The redshift and the references of the compiled 
sample (given below) are presented in the last columns, respectively. The redshift values were taken from the NASA/IPAC Extragalactic Database (NED).
Full table is available online.}
\label{tab1}
\begin{tabular}{lccccccccccccc}	 
\noalign{\smallskip} 
\hline 
                                    &    \multicolumn{2}{c}{[O\,II]$\lambda \lambda$3726,29}   &                                    &    \multicolumn{2}{c}{[O\,III]$\lambda$5007 }   &            &      \multicolumn{2}{c}{ [N\,II]$\lambda$6584}     &                        &      \multicolumn{2}{c}{ [S\,II]$\lambda \lambda$6716+31} & redshift   & Ref.  \\
\cline{2-3}
\cline{5-6}
\cline{8-9}
\cline{11-12}
Object                         &         Obs.                 &             Mod.                        &                                   &           Obs.                     &           Mod.                 &            &             Obs.                       &              Mod.               &                             &         Obs.                 &  Mod.       &            &           \\
\hline
IZw\,92                        &           2.63               &       2.65                              &                                   &         10.12                      &          9.19                  &             &         0.97                          &          1.01                    &                            &        0.77                   &      0.74       &  0.0378        &       1    \\                                
NGC\,3393                   &          2.41                &         2.61                            &                                   &         16.42                      &      13.15                     &             &         4.50                          &            4.42                  &                            &        1.53                   &       1.37  &  0.0125        &        2   \\
\hline 
\end{tabular}
\end{table*}

\begin{table*}\small
\caption{ Assumed model parameter values used to fit the emission lines observed in the 44 modeled Seyfert 2 nuclei. Full table is available online.}
\label{tab2}
\begin{tabular}{lcccccc}	 
\noalign{\smallskip} 
\hline 
Object                     &$\rm \log(O/H)$ &    $\rm \log(N/H)$                & $\rm \log(S/H)$       &  $N_{\rm e}$ (cm$^{_3}$) &      log[Q(H)]  & $\alpha$            \\
\hline
IZw\,92                    &      $-$3.4256       &        $-$4.1850                  &       $-$4.9190       &      822          &       51.27       &        $-$1.4           \\                                
NGC\,3393                  &      $-$3.0583       &       $-$3.4935                   &      $-$4.3300        &      4162         &      50.55        &        $-$1.0          \\
\hline 
\end{tabular}
\end{table*}

From the 61 objects  in our sample, it was possible to obtain  detailed 
 photoionization model solutions for  44 of them.  
The predicted emission line intensities and the final model parameters  for these 44 objects are
listed in Table~\ref{tab1} and \ref{tab2}, respectively. 

In Fig.\ \ref{fres}, bottom  part of each panel,  the observational emission line intensities  (with respect to H$\beta$) are compared 
with those predicted by the detailed photoionization models, for each of the 44 fitted object of the sample.  Also in Fig.\ \ref{fres}, top    part of each panel,
the ratio between  the observed and the modelled intensities as a function of the observed ones is shown.
 We can note that, for most cases, the larger discrepancy is derived for the [\ion{O}{iii}]$\lambda$5007/H$\beta$ ratio, where 
 the intensities observed are, in general,  higher than those predicted by the models. 
  This  could be due to  the presence of a  secondary
source of ionization and   heating in  the   AGNs  (not considered in the photoionization models)  producing  a 
more pronounced increase in the [\ion{O}{iii}] line intensities;  hence, these emission lines  have a strong dependence with the electron temperature
  (see also \citealt{dors15, enrique10}).  
Moreover, the supposition of  electron density constant along the radius of the narrow
line region, do not treat all the relevant physical processes correctly  (e.g. process of gas cooling  by free-free
and free-bound emission,  or temperature fluctuations), and  the used of simplified geometry in the models,
could also be responsible  for this discrepancy.

\begin{figure}
\centering
\includegraphics[angle=-90,width=0.8\columnwidth]{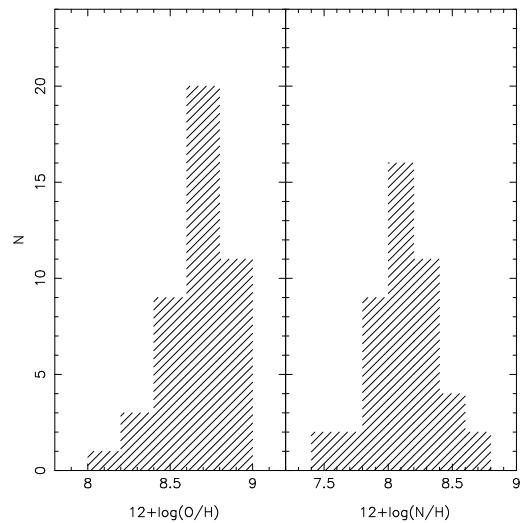}
\caption{Histogram containing the  oxygen (left panel)  and nitrogen (right panel)  abundance distributions predicted by our photoionization models 
for the  44 fitted objects of our sample (see Sect.~\ref{resdisc}). }
\label{fh}
\end{figure}

In Fig.~\ref{fh},  we present a  histogram containing the   oxygen and nitrogen abundances relative to the
hydrogen abundance,  predicted by our photoionization models  for  the   44 fitted objects of our sample (listed in Table~\ref{tab2}). 
 Analysing the  obtained oxygen abundance distribution, we found   that  $\sim70\%$ of the
 objects present  O/H abundances in the range $\rm 8.6 \: \la 12+log(O/H) \: \la \: 9.0$  (or metallicities
 in the range  $ 1.0 \: \la (Z/Z_{\odot}) \: \la \: 2.0$). This result is in consonance with the  results
 found by   \citet{castro17} and \citet{dors15},  who considered a different methodology to 
 estimate the metallicity for a similar sample of objects.
 
 Regarding the nitrogen  abundance distribution also presented in Fig.~\ref{fh}, we found that 
 Sy2 nuclei exhibit a wide  range of N/H abundances, i.e.  $\rm 7.4 \: \la \: 12+log(N/H) \: \la \: 8.8$ or,  
   $\rm  0.3 \: \la (N/N_{\odot}) \: \la \: 7.4$. This is a   wider range of values
 than the one derived by \citet{thaisa90}. The discrepancy between these estimations  is probably
  due to  differences in the studied sample or due to differences in the employed methodology.

\begin{figure} 
\centering
\includegraphics[angle=-90,width=0.8\columnwidth]{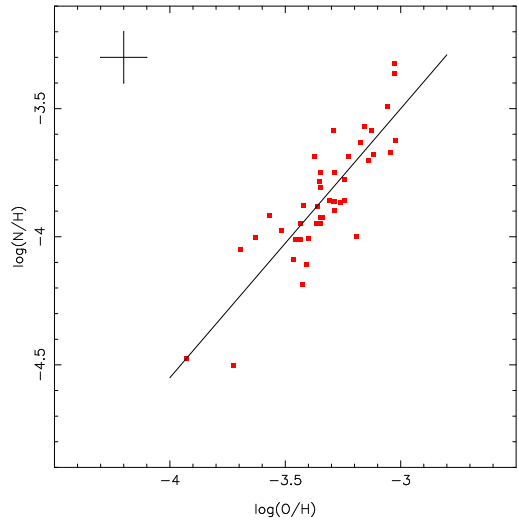}
\caption{$\rm \log(N/H)$  vs.  $\rm log(O/H)$ predicted  by the models. The line represents a linear regression
 fitting the points, given by $\rm \log(N/H)=(1.05 \pm0.09) \times [\log(O/H)] -(0.35 \pm 0.33$). 
 Error bars represents the uncertainty 0.2 dex   in our abundance estimations (see Sect.~\ref{mod2}).}
\label{fa}
\end{figure}

Our abundance results make possible to derive a relation between nitrogen and oxygen abundances in  Sy2s. With this goal,  in Fig.~\ref{fa},
we  show log(N/H) versus  log(O/H) (listed in Table~\ref{tab2}). 
 We  see a clear correlation indicating that the nitrogen has a secondary origin 
for $\rm 12+log(O/H) \: > \:8.0$, in Sy2 nuclei. A linear regression  fitting these data produce: 
\begin{eqnarray*}
\rm 
\log(N/H) = (1.05 \pm0.09) \times [\log(O/H)] -(0.35 \pm 0.33).
\end{eqnarray*}

\begin{figure}
\centering
\includegraphics[angle=-90,width=0.8\columnwidth]{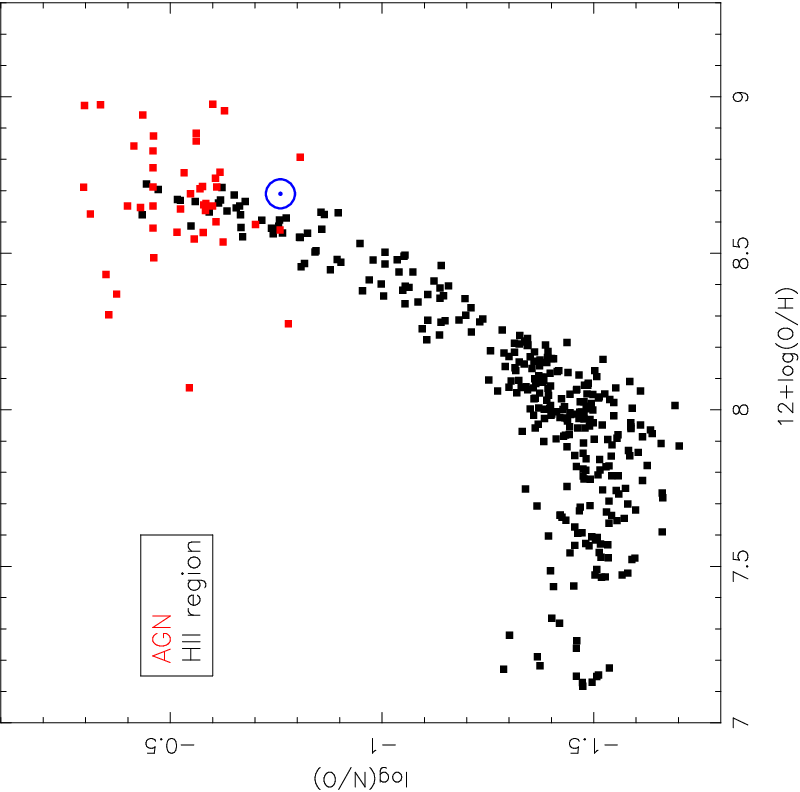}
\caption{log(N/O)  vs.  12+log(O/H) abundance ratio values. The red points are values predicted by the individual photoionization models for our sample of
Sy2 objects (see Sect.~\ref{resdisc}), while the 
 black points are estimations for  \ion{H}{ii} regions derived by \citet{pily16} 
   using  the  C-method  \citep{pilyugin12}. The value for the solar ratio abundance taken from
\citet{holweger01} and  \citet{grevesse98} is also indicated in the plot with a blue symbol.}
\label{fb}
\end{figure}

 In Fig.~\ref{fb},  the predicted N/O  as a function of O/H values for the AGN sample is compared with those derived for a sample of \ion{H}{ii} 
 regions by \citet{pily16}, who used the  C-method \citep{pilyugin12}  to calculate the oxygen and nitrogen abundances.  In this plot, the value for the solar 
abundance is  also indicated. We can see that the N/O estimations for the Sy2 objects  are in agreement with those for the metal richest \ion{H}{ii} 
regions. Moreover, the N/O values of the objects of our sample are, in most cases, higher than the solar value.

\section{Conclusion}
\label{conc}

We compiled from the literature narrow optical emission line intensities for
a sample of  Seyfert 2  galaxies. Standard  photoionization  models
were built in order to reproduce the intensities  of these emission lines.
 We present new results of   nitrogen and oxygen abundances calculated  for a large sample of Seyfert 2
galaxies.   We found a very wide range for the N/H abundances in this kind of active nuclei, varying between about 0.3 and 7.5 times the solar value. 
We derived a relationship between the nitrogen and oxygen abundances for Seyfert 2 AGNs.  
 We find that N/O abundance ratios in Seyfert 2 galaxies are similar to those
recently derived for a sample of extragalactic \ion{H}{ii} regions with high metallicity.
   
\section*{Acknowledgments}
We are grateful to the referee, Dra  Marcella Contini, for her useful
comments and suggestions, which have helped us to substantially
clarify and improve the manuscript. O.L.D. is grateful to  FAPESP (2016/04728-7)
 and CNPQ (306744/2014-7).

\label{lastpage}

\end{document}